# Level spacing of U(5) ↔ SO(6) transitional region with maximum likelihood estimation method

M. A. Jafarizadeh[a,b] [*], N. Fouladi[c][†], H. Sabri[c], B. Rashidian Maleki[c]

[a]Department of Theoretical Physics and Astrophysics, University of Tabriz, Tabriz 51664, Iran.

[b]Research Institute for Fundamental Sciences, Tabriz 51664, Iran.

[c]Department of Nuclear Physics, University of Tabriz, Tabriz 51664, Iran.

[*] E-mail: jafarizadeh@tabrizu.ac.ir
[†] E-mail: fouladi@tabrizu.ac.ir

# Abstract

In this paper, a systematic study of quantum phase transition within $U(5) \leftrightarrow SO(6)$ limits is presented in terms of infinite dimensional Algebraic technique in the IBM framework. Energy level statistics are investigated with Maximum Likelihood Estimation (MLE) method in order to characterize transitional region. Eigenvalues of these systems are obtained by solving Bethe-Ansatz equations with least square fitting processes to experimental data to obtain constants of Hamiltonian. Our obtained results verify the dependence of Nearest Neighbor Spacing Distribution's (NNSD) parameter to control parameter ($c_s$) and also display chaotic behavior of transitional regions in comparing with both limits. In order to compare our results for two limits with both GUE and GOE ensembles, we have suggested a new NNSD distribution and have obtained better KLD distances for the new distribution in compared with others in both limits. Also in the case of N→∞, the total boson number dependence displays the universality behavior, namely NNSD tends to Poisson limit for every values of control parameter.

# Introduction

Investigation of transitional behavior between dynamical symmetry limits has become to an interesting topic in recent years [1-20]. Level crossing [3,7-10], significant variation in the intensities of electromagnetic transitions [3-15] and etc can be used to characterize these regions in different nuclei. On the other hand, dynamical symmetry means the integrability of the system in classical limit and constants of motion associated with a symmetry govern the integrability of the system or regular behavior in these limits in compared to transitional regions where a mixed symmetry visualize with nuclei[13-20]. In this point of view, statistical properties of nuclear spectra can be used as new characteristic to clarify transition, which one can predict a chaotic dynamic for transitional region in comparing with regular ones for symmetry limits. Also, an explicit relation between control parameter of any transitional Hamiltonian and statistical behavior of nuclei can verify our prediction about this new criterion for transition region. In order to study transitional systems in Interaction Boson Model (IBM)[21-26], U(6) Lie Algebra must be used. These methods normally require to diagonalize the Hamiltonian by using of numerical methods which lead to some unexpected uncertainty for results [3,4,27]. In order to simplify this method, an affine Lie algebra su(1,1) without central extension approach was suggested[27-28] which evaluates experimental spectra similar to Hamiltonian in geometric collective model framework[31-32].

On the other hand, the statistical properties of nuclear spectra, can be investigated with different methods in which the Nearest Neighbor Spacing Distribution (NNSD) or P(s) functions are suggested as the best ones. In usual methods [33-44], one can apply least square fitting processes to every sequence (distribution of level spacing after unfolding processes) with well-known distributions as Brody distribution [38] and etc. The value of every distribution's parameter characterizes chaotic (Wigner limit) or regular (Poisson limit) behavior but the results of this procedure has some unusual uncertainty and also it can't lead to acceptable results in cases with small size of data .We have suggested a new method [45] (which is based on Maximum Likelihood Estimation (MLE) method) to estimate every distribution's parameter which as [46] (Bayesian method for estimation) can lead to very exact results with low uncertainty.(our results are very close to Cramer-Rao Lower Bound(CRLB))[47].

Also in order to complete our analysis about the similarity of nuclear spectra with Gaussian Unitary Ensembles(GUE) and Gaussian Orthogonal Ensembles(GOE), we have suggested a new distribution for NNSDs which describes all Poisson, GOE and GUE limits.(other distribution, only explain two limits of these limits and this new distribution can describe systems in general case). Also, we have applied the MLE method [45] to investigate this new proposed distribution.

In section 3, we have reviewed an algebraic Beth-Ansatz method to diagonalizes the $su(1,1)$ transitional Hamiltonian. In order to investigate statistical behavior of transitional region, we applied this procedure to systems with total boson number N = 8,9,10. Firstly, we have calculated constant of Hamiltonian($h^k$) with Beth-Ansatz method, then in order to obtain constants of eigenvalues, we have evaluated all energy levels that construct our used sequences with least square fitting to experimental data[48-54](experimental spectra of nuclei with these boson numbers which visualize these symmetries such as $^{116}_{52}Te$ , $^{150}_{64}Gd$ , $^{152}_{66}Dy$ , ...for U(5) limit[3,48], $^{192}_{80}Hg$ , $^{130}_{58}Ce$ , $^{188}_{80}Hg$, ... for SO(6) limit[3,48,49] and Ru-Pd [48-49] and also Xe, Ba [1-3] regions for transitional regime). By applying the above mentioned method (MLE method to estimate Brody and our new suggested distribution parameters), results show chaotic behavior for transitional region ($c_s$: 0.4~0.6) in compared to regular dynamics for both symmetry limits.

We also display this chaotic behavior for transitional region with experimental data (sequences which have prepared from all nuclei used in previous part). Also with Kullback-Leibler Divergence (KLD), we have evaluated the distances of our new suggested distribution to both GUE and GOE. The obtained results, confirm the better distances of our distribution to both limits per values obtained by MLE method in compared with other ones.  Also this closer approach of our results in special sequences constructed by O(6) nuclei in compared with other sequences, confirm theoretical prediction[37-45] about chaotic behavior of this symmetry limit related to U(5) symmetry. We have also evaluated CRLBs of Brody and our new distributions and results display the smallest bound for this new distribution in compared to others in the same sequence. We also apply this method to cases with N=25,50,100, and confirm previous results[13-14] about universality behavior (tend to Poisson limit) for these systems in the case $N \to \infty$.

This paper is organized as follows: In section 2, an affine $su(1,1)$ Algebra and statistical approaches is presented which will be used for investigating the statistical properties of transitional regions and also a new distribution is suggested for investigating the statistical properties in general form. In section 3, the numerical results about statistical behaviors of different systems in transition region and these both symmetry limits would be presented. We also will represent some conclusions about dependence of control parameter to chaoticity of systems ,closer distance of new distribution to both limits and  also universality behavior of distribution function in limit $N \to \infty$.

## 2. The affine su(1,1) based Hamiltonian  and  statistical formalism

### a) su(1,1) approach to transitional region

The $SU(1,1)$ Algebra have been described in detail in Ref[1-4,14-29].So we only mention on the main results which have been analyzed in this article. The Lie algebra corresponding to the group $SU(1,1)$ is spanned by the three operators $\{S_1, S_2, S_0\}$,

$[S_1, S_2] = -iS_0 \qquad , \qquad [S_2, S_0] = iS_1 \qquad , \qquad [S_0, S_1] = iS_2 \qquad (1a)$

It is convenient to use raising and lowering operators $S_\pm = S_1 \pm iS_2$ which satisfy the following commutation relations:

$[S^0, S^\pm] = \pm S^\pm \qquad , \quad [S^+, S^-] = -2S^0 , \qquad (1b)$

The Casimir operator of this Algebra can be introduced as follow:

$$\hat{C}_2 = S^0(S^0 - 1) - S^+S^-, \tag{2}$$

Representations of $SU(1,1)$ are determined by a single number $\kappa$, thus the representation of Hilbert space is spanned by the orthonormal basis $|\kappa\mu\rangle$ where $\kappa$ can be any positive number and $\mu = \kappa, \kappa+1, \ldots$, then we can write

$$\hat{C}_2(SU(1,1))|\kappa\mu\rangle = \kappa(\kappa-1)|\kappa\mu\rangle, \qquad S^0|\kappa\mu\rangle = \mu|\kappa\mu\rangle, \tag{3}$$

Now we can introduce the infinite dimensional algebra that is generated by using of

$$S_n^\pm = c_s^{2n+1} S^\pm(s) + c_d^{2n+1} S^\pm(d) \quad , \quad S_n^0 = c_s^{2n} S^0(s) + c_d^{2n} S^0(d) \tag{4}$$

$c_s$ and $c_d$ are real parameters and $n$ can be $0, \pm 1, \pm 2, \ldots$. These generators satisfy the commutation relations

$$[S_m^0, S_n^\pm] = \pm S_{m+n}^\pm \quad , \quad [S_m^+, S_n^-] = -2 S_{m+n+1}^0 \tag{5}$$

Then, the $\{S_m^\mu, \mu = 0, +, -; m = 0, \pm 1, \pm 2, \ldots\}$ makes an affine Lie algebra $\widehat{SU(1,1)}$ without central extension. Now, we can utilize generators of $\widehat{SU(1,1)}$ Algebra to introduce the following Hamiltonian for transitional region between $SO(6) \leftrightarrow U(5)$ limits [25-29]

$$\hat{H} = g S_0^+ S_0^- + \alpha S_1^0 + \gamma \hat{C}_2(SO(5)) + \delta \hat{C}_2(SO(3)) . \tag{6}$$

$g, \alpha, \gamma$ and $\delta$ are real parameters. It can be shown that (6), would be equivalent with the SO(6) Hamiltonian if $c_s = c_d$, and with U(5) Hamiltonian when $c_s = 0$ & $c_d \neq 0$. Therefore, the $c_s \neq c_d \neq 0$ requirement just correspond to the $SO(6) \leftrightarrow U(5)$ transitional region. In our calculation, we take $c_d$ (=1) constant value and vary $c_s$ between 0 and $c_d$ [29-32].

For evaluating the eigenvalues of Hamiltonian (4), let us to write eigenstates of (4) (as mentioned in [29-32])

$$|k; v_s v n_\Delta LM\rangle = \sum_{n_i \in Z} a_{n_1 n_2 \ldots n_k} x_1^{n_1} x_2^{n_2} \ldots x_k^{n_k} S_{n_1}^+ S_{n_2}^+ \ldots S_{n_k}^+ |lw\rangle, \tag{7}$$

Because of the analytical behavior of the wavefunctions, it suffices to consider $x_i$ near zero. Now if we take commutation relations between generators of $SU(1,1)$ Algebra (5), wavefunctions express as (more details about these concepts are presented in [29-35])

$$|k; v_s v n_\Delta LM\rangle = \mathcal{N} S_{x_1}^+ S_{x_2}^+ \ldots S_{x_k}^+ |lw\rangle, \tag{8}$$

where $\mathcal{N}$ is the normalization factor and

$$S^+_{x_i} = \frac{c_s}{1 - c_s^2 x_i} S^+(s) + \frac{c_d}{1 - c_d^2 x_i} S^+(d), \tag{9}$$

The c-numbers $x_i$ are determined by the following set of equations

$$\frac{\alpha}{x_i} = \frac{g c_s^2 \left(v_s + \frac{1}{2}\right)}{1 - c_s^2 x_i} + \frac{g c_d^2 \left(v + \frac{5}{2}\right)}{1 - c_d^2 x_i} - \sum_{j \neq i} \frac{2}{x_i - x_j} \quad for\ i = 1,2,\dots,k\ , \tag{10}$$

The eigenvalues $E^{(k)}$ of Hamiltonian (4) can then be expressed

$$h^{(k)} = \sum_{i=1}^{k} \frac{\alpha}{x_i}, \tag{11}$$

which

$$E^{(k)} = h^{(k)} + \gamma v(v+3) + \delta L(L+1) + \alpha \Lambda_1^0, \tag{12}$$

and

$$\Lambda_1^0 = \frac{1}{2}\left[c_s^2\left(v_s + \frac{1}{2}\right) + c_d^2\left(v + \frac{5}{2}\right)\right], \tag{13}$$

The quantum number ($k$) is related to total boson number $N$ by

$$N = 2k + v_s + v$$

## b) Statistical analysis of nuclear spectra

The statistical analysis of nuclear spectra can be studied by different methods. All of them (such as Nearest Neighbor Spacing Distribution (NNSD) [33-38], the Dyson-Mehta $\Delta_3$ statistic [34] and etc) have been carried with comparison of fluctuation properties of selected spectrum with theoretical predictions of Random Matrix Theory (RMT), integrable (ordered) systems or interpolation between these two chaotic and regular limits. In NNSD method (we have restricted our analysis to this method), level spacing of nuclear spectra have been prepared with unfolding processes to compare with theoretical accounts The distribution $P(s)$ is the best spectral statistic to analyze shorter series of energy levels and the intermediate regions between order and chaos. To unfold our spectrum, we must use some levels with same symmetry [35-37]. This requirement means to use levels with same total quantum number (J) and same parity which these collection of levels will be called "sequence" [35-40] (in some sequences, we have used all $2^+, 4^+, 6^+$ levels because small size of data don't allow exact analysis). Then we first include the number of the levels below $E$ and write it as

$$N(E) = N_{ave}(E) + N_{fluc}(E)$$

Then with taking a smooth polynomial function of degree 6 to fit the staircase function, we fix $N_{ave}(E)$. Therefore, the unfolded spectrum with the mapping $E_i \rightarrow \epsilon_i$

$$\epsilon_i = N_{ave}(E_i)$$

The nearest-neighbor level spacing is defined as $s_i (\equiv \epsilon_{i+1} - \epsilon_i)$ which unfolded sequence $\{s_i\}$ is clearly dimensionless and has a constant average spacing of 1, then distribution $P(s)$ will be as $P(s)ds$ that is the probability for the $s_i$ to lie within the infinitesimal interval [s,s+ds]. It has been shown that the nearest-neighbor spacing distribution $P(s)$ measures the level repulsion. For sequences with properties similar to GOE statistics, NNSD probability distribution function is approximated with Wigner distribution [33-35]

$$P(s) = \frac{1}{2}\pi s e^{-\frac{\pi s^2}{4}}, \qquad (14)$$

Investigation of the statistical data from proton and neutron resonance for different nuclei that demonstrates the NNSD for levels with excitation energy about 8Mev(particle emission threshold) is well represented by a Wigner distribution. On the other hand, one can similarly show that in non-interacting systems (where a number of vanishing H-matrix elements appear because of the presence of certain symmetries, for example isospin symmetry that govern Hamiltonian describing the system), the energy spacing is described by Poisson distribution [33-34]

$$P(s) = e^{-s}, \qquad (15)$$

Interpolation between these two limits for statistical behavior of different systems have displayed with results of different groups [35-37] (these results verify theoretical predictions about mixture of regular and chaotic dynamics for low-lying energy levels of excited nuclei [33-34]). . In order to quantify the chaoticity of $P(s)$ in terms of a parameter, it can be compared for example to the Brody or the Berry–Robnik distributions, which are adequate for description of intermediate situations between order and chaos. Although each of these distributions has some advantage in limiting cases, they are very similar in a particular case like ours. We use here the Brody distribution [38], given by

$$P(s) = b(1+q)s^q e^{-bs^{q+1}} \qquad b = \left[\Gamma\left(\frac{2+q}{1+q}\right)\right]^{q+1}, \qquad (16)$$

Which consider a power-law level repulsion and interpolates between the Poisson (q = 0) and Wigner (q = 1) distributions. This distribution cannot explain another limit (GUE which will use in describing another symmetry limit). In order to investigate all limits (Poisson, GUE and GOE), we will introduce a new distribution in the following.

- **New distribution**

The phenomenon of level repulsion in nuclear energy spectra has been investigated in different papers [33-40] and several distributions has been suggested in order to describe behavior of system between Poisson (order limit) and three limits of Random Matrix Theory (RMT) namely GOE,GUE and GSE [38,43,46]. All of these distributions, only describe interpolation between Poisson and one of these three limits and cannot display a general behavior for all ones. In order to introduce a new distribution which investigate Poisson (order),GOE(Wigner or chaotic) and GUE limits, we have proposed another statistics derived from Wigner surmise. The nearest neighbor spacing of Gaussian orthogonal ensemble was distributed as (14). On the other hand, the nearest neighbor spacing of Gaussian unitary ensemble can be described by

$$P(s) = \frac{32}{\pi^2} s^2 e^{-\frac{4s^2}{\pi}}. \tag{17}$$

We extended both (14) and (17) relations by means of ansatz

$$P(s) = b(1+q)(\alpha s^q + \beta s^{q+1})e^{-bs^{q+1}}, \tag{18}$$

With applying the normalization requirements

$$\int_0^\infty P(s)\, ds = 1 \qquad \& \qquad \int_0^\infty s\, P(s)\, ds = 1$$

We can obtain the constants of (18) as

$$\alpha = 1 - \frac{\left(\frac{\Gamma\left[\frac{q+2}{q+1}\right]}{b^{\frac{1}{1+q}}}\right)^2 - \frac{\Gamma\left[\frac{q+2}{q+1}\right]}{b^{\frac{1}{1+q}}}}{\left(\frac{\Gamma\left[\frac{q+2}{q+1}\right]}{b^{\frac{1}{1+q}}}\right)^2 - \frac{\Gamma\left[\frac{q+3}{q+1}\right]}{b^{\frac{2}{1+q}}}} \quad , \quad \beta = \frac{\left(\frac{\Gamma\left[\frac{q+2}{q+1}\right]}{b^{\frac{1}{1+q}}}\right) - 1}{\left(\frac{\Gamma\left[\frac{q+2}{q+1}\right]}{b^{\frac{1}{1+q}}}\right)^2 - \frac{\Gamma\left[\frac{q+3}{q+1}\right]}{b^{\frac{2}{1+q}}}} \tag{19}$$

In the following, we will apply the MLE method [45] to this new distribution and display estimators for all parameters.

- **Estimation the parameter of Brody's distribution with MLE**

As mentioned in [16-20,33- 40], the small size of data cause to unusual uncertainty for results which obtained from Least Square fitting processes. The MLE method provides an opportunity for obtaining exact result with minimum variation (the results are closer to Cramer-Rao Lower Bound (CRLB)) .Here we propose a generalized Brody distribution with two parameters of b and q as:

$$P(s) = b(1+q)s^q e^{-bs^{q+1}},$$

Where it reduces to Brody one by choosing $b = \left[\Gamma\left(\frac{2+q}{1+q}\right)\right]^{q+1}$.

Now, we must choose the adequate maximum likelihood estimators to estimate the parameters b and q. For this purpose, we try to use the products of the generalized Brody distribution functions as a likelihood function [45,47], namely:

$$L(q,b) = \prod_{i=1}^{n} b(1+q)s_i^q e^{-bs_i^{q+1}} = [b(1+q)]^n \prod_{i=1}^{n} s_i^q \, e^{-b\sum s_i^{q+1}}, \quad (20)$$

Then, we obtain the following pair of implicit equations for the required estimators by derivation of the logarithm of likelihood function (20) with respect to the parameters and setting them to zero, i.e., maximizing likelihood function:

$$f_1: \frac{1}{n}\sum s_i^{q+1} - \frac{1}{b}, \qquad \text{for } b \qquad (21)$$

$$f_2: \frac{b}{n}\sum \ln s_i \, s_i^{q+1} - \frac{1}{n}\sum \ln s_i - \frac{1}{1+q}, \qquad \text{for } q \qquad (22)$$

Now, the parameters b and q can be estimated by very accurate solving of above equation through Newton-Raphson iteration method which finally tend to the following relations [45]

$$q_{new} = q_{old} - \frac{\left[\frac{1}{n}\sum \ln s_i \, s_i^{q+1}\right]\left[\frac{1}{n}\sum s_i^{q+1} - \frac{1}{b}\right] - \frac{1}{b^2}\left[\frac{b}{n}\sum \ln s_i \, s_i^{q+1} - \frac{1}{n}\sum \ln s_i - \frac{1}{1+q}\right]}{\left[\frac{1}{n}\sum \ln s_i \, s_i^{q+1}\right]^2 - \frac{1}{b^2}\left[\frac{b}{n}\sum (\ln s_i)^2 \, s_i^{q+1} + \frac{1}{(1+q)^2}\right]}\Bigg|_{\substack{b \to b_{old} \\ q \to q_{old}}} \quad (23)$$

$$b_{new} = b_{old} - \qquad (24)$$

$$-\frac{\left[-\frac{b}{n}\sum(\ln s_i)^2 \, s_i^{q+1} - \frac{1}{(1+q)^2}\right]\left[\frac{1}{n}\sum s_i^{q+1} - \frac{1}{b}\right] + \left[\frac{1}{n}\sum \ln s_i \, s_i^{q+1}\right]\left[\frac{b}{n}\sum \ln s_i \, s_i^{q+1} - \frac{1}{n}\sum \ln s_i - \frac{1}{1+q}\right]}{\left[\frac{1}{n}\sum \ln s_i \, s_i^{q+1}\right]^2 - \frac{1}{b^2}\left[\frac{b}{n}\sum(\ln s_i)^2 \, s_i^{q+1} + \frac{1}{(1+q)^2}\right]}\Bigg|_{\substack{b \to b_{old} \\ q \to q_{old}}}$$

With iteration processes (with arbitrary times) we can calculate the final values of "b" and "q" with minimum uncertainty. As mentioned in introduction section, the results of MLE are closer to Cramer-Rao Lower Bound (CRLB) which can be displayed with relation

$$CRLB \equiv \frac{1}{MF(q)}\Bigg|_{for\ final\ value\ of\ "q"\ obtained\ from\ MLE}, \qquad (25)$$

$F(q)$ indicates Fisher information [45-47].(Details about application of this relation can be found in [45]).

- **MLE method for new distribution**

As have been introduced in [45] and applied for Brody distribution, we must construct appropriate likelihood function in order to estimate α,β, q and b. With replaying above mentioned process, we have choice likelihood function as a multiplication for all variables as

$$L(q,b,\alpha,\beta) = \prod_{i=1}^{n} b(1+q)(\alpha s_i^q + \beta s_i^{q+1})e^{-bs_i^{q+1}}, \tag{26a}$$

Then, with putting zero all derivates of logarithm (26a) with respect to all variables, we can introduce our used estimators as

$$L(q,b,\alpha,\beta) = (b(1+q))^n \prod_{i=1}^{n}(\alpha s_i^q + \beta s_i^{q+1})e^{-bs_i^{q+1}}, \tag{26b}$$

$$\ln L(q,b,\alpha,\beta) = n\ln(b(1+q)) + \sum_{i=1}^{n}\ln(\alpha s_i^q + \beta s_i^{q+1}) - b\sum_{i=1}^{n} s_i^{q+1}, \tag{26c}$$

Therefore, we can introduce our estimators as

$$\frac{\partial \ln L(q,b,\alpha,\beta)}{\partial q} = 0 \quad \Rightarrow \quad f_1: \frac{n}{1+q} + \sum_{i=1}^{n}\ln s_i - b\sum_{i=1}^{n}\ln s_i\, s_i^{q+1} \qquad for\ q \tag{27a}$$

$$\frac{\partial \ln L(q,b,\alpha,\beta)}{\partial b} = 0 \quad \Rightarrow \quad f_2: \frac{n}{b} - \sum_{i=1}^{n} s_i^{q+1} \qquad for\ b \tag{27b}$$

$$\frac{\partial \ln L(q,b,\alpha,\beta)}{\partial \alpha} = 0 \quad \Rightarrow \quad f_3: \sum_{i=1}^{n}\frac{s_i^q}{\alpha s_i^q + \beta s_i^{q+1}} \qquad for\ \alpha \tag{27c}$$

$$\frac{\partial \ln L(q,b,\alpha,\beta)}{\partial \beta} = 0 \quad \Rightarrow \quad f_4: \sum_{i=1}^{n}\frac{s_i^{q+1}}{\alpha s_i^q + \beta s_i^{q+1}} \qquad for\ \beta \tag{27d}$$

In order to estimate the values of above parameters, we will apply Newton-Raphson iteration method (as introduced in [45,47]).All details about these calculations and final results and also CRLB for this new distribution are presented in Appendix(I) and (II).

## 3. Numerical result

To obtain numerical results for $E^{(k)}$, we must solve a set of non-linear Beth-Ansatz equations (BAE) with k-unknowns for k-pair excitation[27-30]. Now let us, change variables as

$$\beta = \frac{\alpha}{g} \quad (g = 1 \, kev[28-30]) \qquad c = \frac{c_s}{c_d} \leq 1 \qquad y_i = c_d^2 x_i$$

Therefore, the new form of (9) would be [28-30]

$$\frac{\beta}{y_i} = \frac{c^2\left(v_s + \frac{1}{2}\right)}{1 - c^2 y_i} + \frac{\left(v + \frac{5}{2}\right)}{1 - y_i} - \sum_{j \neq i} \frac{2}{y_i - y_j} \quad for \; i = 1,2,\ldots,k \qquad (23)$$

In order to evaluate roots of Beth-Ansatz equations (BAE) for energy levels of every nuclei with specified values of $v_s$ and $v$, we have solved equation (23) with definite values of c and **α**, for $i = 1$ and then we can use "Find root" in Maple13 to get all $y_j$'s (we iterate our calculation with different values of these parameters(c and **α**) to evaluate experimental spectra[48-54](after inserting γ and δ) with minimum variation).

$$\sigma = \left(\frac{1}{N_{tot}} \sum_{i,tot} |E_{exp}(i) - E_{cal}(i)|^2\right)^{1/2}$$

($N_{tot}$ is the number of energy levels in the fitting processes). The method for fixing the best set of parameters in the Hamiltonian (γ and δ) includes carrying out a least-square fit procedure of the excitation energies of selected states ($0_1^+, 2_1^+, 4_1^+, 0_2^+, 2_2^+, 4_2^+, 0_3^+, 3_1^+, 2_3^+, 0_4^+, 6_1^+, 2_4^+$ or other levels from selected nuclei) and the two neutron separation energies of all isotopes in each isotopic chain. We applied Least Square fitting to fit these relations with experimental data [48-54]. We have applied this procedure to all nuclei which have used in our analysis and numerical results about their parameters displayed in below tables.

To study transitional region, we have choose some nuclei which visualize two symmetry limits (U(5) ,SO(6)) and transitional region with total boson numbers N=8,9,10 [3,8,10,49-54]. All constants of spectra related to these nuclei were evaluated with method introduced in previous part with $c_s$ values which was varied between 0 (U(5) limit) and 1 (SO(6) limit). Our obtained results about Hamiltonian's parameters related to every nucleus are presented in tables (1-3).

| Nuclei | $^{116}_{52}Te$ | $^{120}_{52}Te$ | $^{108}_{46}Pd$ | $^{124}_{54}Xe$ | $^{104}_{44}Ru$ | $^{192}_{78}Pt$ | $^{192}_{80}Hg$ |
|---|---|---|---|---|---|---|---|
| α | 600 | 620 | 610 | 615 | 625 | 600 | 615 |
| $C_s$ | 0 | 0.15 | 0.27 | 0.42 | 0.64 | 0.83 | 1 |
| γ | 95.42 | 94.32 | 60.88 | 72.91 | 75.34 | 65.91 | 148.23 |
| δ | 11.07 | 11.08 | 11.75 | -0.7280 | 21.18 | -2.01 | -31.09 |
| σ | 101 | 123 | 89 | 139 | 117 | 95 | 104 |

Table1.Parameters of energy spectra for different nuclei with total boson number N=8.units of $\alpha, \delta, \gamma$ are in kev.

| Nuclei | $^{150}_{64}Gd$ | $^{114}_{48}Cd$ | $^{110}_{46}Pd$ | $^{106}_{44}Ru$ | $^{122}_{54}Xe$ | $^{190}_{80}Hg$ | $^{130}_{58}Ce$ |
|---|---|---|---|---|---|---|---|
| $\alpha$ | 600 | 610 | 620 | 615 | 600 | 610 | 615 |
| $C_s$ | 0 | 0.19 | 0.46 | 0.53 | 0.57 | 0.81 | 1 |
| $\gamma$ | 60.68 | 58.78 | 40.38 | 64.47 | 35.48 | 65.35 | 59.55 |
| $\delta$ | 4.53 | 6.37 | 14.94 | -4.11 | 6.28 | 1.82 | -3.42 |
| $\sigma$ | 84 | 79 | 101 | 94 | 123 | 107 | 117 |

Table2. Parameters of energy spectra for different nuclei with total boson number N=9. units of $\alpha, \delta, \gamma$ are in kev.

| Nuclei | $^{152}_{66}Dy$ | $^{152}_{64}Gd$ | $^{108}_{44}Ru$ | $^{112}_{46}Pd$ | $^{120}_{54}Xe$ | $^{190}_{78}Pt$ | $^{188}_{80}Hg$ |
|---|---|---|---|---|---|---|---|
| $\alpha$ | 600 | 620 | 600 | 610 | 615 | 610 | 620 |
| $C_s$ | 0 | 0.14 | 0.32 | 0.49 | 0.59 | 0.80 | 1 |
| $\gamma$ | 92.94 | 60.44 | 33.61 | 32.81 | 49.17 | 24.23 | 69.05 |
| $\delta$ | 0.02 | -5.13 | 6.73 | 15.58 | 2.45 | 12.22 | 0.43 |
| $\sigma$ | 125 | 141 | 102 | 98 | 133 | 114 | 127 |

Table3. Parameters of energy spectra for different nuclei with total boson number N=10. units of $\alpha, \delta, \gamma$ are in kev.

By using of these values, we evaluated all $2^+, 4^+, 6^+$ levels of these nuclei below $\leq 7 Mev$. Then our used sequence will be constructed by unfolding processes and the Brody distribution's parameter would be obtained by using of the relations (23-24). As have shown in table(4), variation of "q" in all cases, display a dependence to $c_s$, so the chaotic behavior of nuclei increases in the region $c_s: 0 \to \sim 0.5$ while the maximum values of Brody distribution's parameter will be for $c_s: 0.4 \to 0.6$. On the other hand, when $c_s$ increases from this region to the other symmetry limit ($c_s = 1$ or SO(6) limit), q's values decrease and tend to definite values which presented in [43,45] (the results in[45] for O(6) have been displayed with experimental data for this symmetry).

| q | $^{116}_{52}Te$ | $^{120}_{52}Te$ | $^{108}_{46}Pd$ | $^{124}_{54}Xe$ | $^{104}_{44}Ru$ | $^{192}_{78}Pt$ | $^{192}_{80}Hg$ |
|---|---|---|---|---|---|---|---|
|   | 0.45 ± 0.05 | 0.57 ± 0.08 | 0.65 ± 0.06 | 0.67 ± 0.09 | 0.72 ± 0.07 | 0.58 ± 0.11 | 0.54 ± 0.06 |
| q | $^{150}_{64}Gd$ | $^{114}_{48}Cd$ | $^{110}_{46}Pd$ | $^{106}_{44}Ru$ | $^{122}_{54}Xe$ | $^{190}_{80}Hg$ | $^{130}_{58}Ce$ |
|   | 0.43 ± 0.08 | 0.54 ± 0.06 | 0.68 ± 0.05 | 0.67 ± 0.09 | 0.60 ± 0.07 | 0.52 ± 0.08 | 0.47 ± 0.05 |
| q | $^{152}_{66}Dy$ | $^{152}_{64}Gd$ | $^{108}_{44}Ru$ | $^{112}_{46}Pd$ | $^{120}_{54}Xe$ | $^{190}_{78}Pt$ | $^{188}_{80}Hg$ |
|   | 0.46 ± 0.05 | 0.52 ± 0.09 | 0.56 ± 0.06 | 0.63 ± 0.05 | 0.69 ± 0.07 | 0.56 ± 0.04 | 0.50 ± 0.07 |

Table4. Brody distribution's parameter for different nuclei with specified total boson number, displaying chaoticity for transitional region in compare with regular behavior of symmetry limits.

Figures 1,2,3 display NNSDs for different nuclei with definite values of $c_s$ in systems with total boson number N=8,9,10 respectively. In figure 4, we have displayed the CRLB's curves for three cases which display tending of our results to this bound. In figure 5, the relation of "q" to $c_s$ for N=8,9,10 have been showed which display the most chaoticity in the region $c_s$: ~0.4 → ~0.6. As it was mentioned in introduction section, we have controlled the relation between "q" and $c_s$ with experimental spectra too. In order to prepare sequences, we have collected all $2^+$ and $4^+$ levels (the small size of every nuclei's levels made impossible the suitable analysis with unique one and we had to construct our used sequence with unfolding processes from all ones) from used nuclei in previous part in three regions: first region ($c_s$: 0 → ~0.2), second one from nuclei with ($c_s$: ~0.8 → ~1) and the third region from nuclei corresponding to($c_s$: ~0.4 → ~0.6). In table (5), the values of "q" verify our previous results from theoretical values and display chaoticity for transitional region in comparing with both symmetry limits. In Figure 6, we have presented three NNSD's for these three regions with experimental data. These behaviors are similar with the Iachello's prediction about dependence of phase transition to control parameter (in [1-25],η is introduced as control parameter for describing phase transition between (η=0) vibrational(U(5)) and (η=1) γ-unstable rotation (O(6))limits and phase transition in the region η~0.5).Therefore we deduce, $c_s$ as control parameter (similar to η ) can be used to describe these two limits and also transitional regions.

| Nuclei | Nuclei with $c_s$: 0 → ~0.2 | Nuclei with $c_s$: ~0.4 → ~0.6 | Nuclei with $c_s$: ~0.8 → 1 |
|---|---|---|---|
| Brody distribution's parameter | 0.44 ± 0.08 | 0.65 ± 0.09 | 0.53 ± 0.06 |

Table5.Brody distribution's parameter for three regions. Used sequences constructed by experimental data [48-54].

As have explained in previous parts, we have suggested the new distribution (18) to analysis statistical properties of used sequence in general form and it's relation to GOE,GUE and Poisson limits. In order to carry out this procedure, we have applied the relations (I-9 to 13) to sequences which have been constructed of nuclei introduced in table (1)(we don't repeat this procedure for other tables(2,3) as they had the same final result ). We also, evaluated Kullback-Leibler Divergence [45,47] (as below) to obtain distance of our results to GOE (chaotic limit).

$$D_{KL}(P\|Q) = \sum_i P(i) \log \frac{P(i)}{Q(i)} \qquad (24)$$

In which it would display closer distances between two distributions if $D_{KL}(P\|Q) \to 0,$. The values of "q" ,"b","α "and "β" which have been evaluated by MLE method and also KLD for every nuclei are tabulated in table(6).

| Nuclei | $^{116}_{52}Te$ | $^{120}_{52}Te$ | $^{108}_{46}Pd$ | $^{124}_{54}Xe$ | $^{104}_{44}Ru$ | $^{192}_{78}Pt$ | $^{192}_{80}Hg$ |
|---|---|---|---|---|---|---|---|
| $b$ | 0.5711 | 0.4001 | 0.5261 | 0.2891 | 0.4050 | 0.7996 | 0.4261 |
| $q$ | 0.4284 | 0.4775 | 0.6574 | 0.5336 | 0.6784 | 0.5945 | 0.4890 |
| $\alpha$ | 1.5071 | 1.8536 | 1.6259 | 2.2400 | 1.9231 | 1.0746 | 1.8045 |
| $\beta$ | -0.3772 | -0.5079 | -0.4757 | -0.6631 | -0.6029 | -0.727 | -.5018 |
| $KLD$ | 1.5004 | 1.2072 | 0.9672 | 0.9047 | 0.7241 | 1.1056 | 1.2544 |

Table6. The parameters of new distribution which evaluated from sequences constructed from nuclei introduced in table (1) and also the KLD related to GOE for every one.

As have displayed in table (6), the values of KLDs verify our previous results about chaotic behavior of transitional regions in compared to both symmetry limits (the small distances to GOE for nuclei in transitional region show this behavior). Also in order to compare our results for this new distribution with Brody distribution's values, we have calculated KLD to obtain distances to GUE and GOE in both U(5) and SO(6) limits for nuclei with total boson number N=8,9,10; these results are displayed in tables(7,8).

| KLD for different distributions | Nuclei with U(5) symmetry | | |
|---|---|---|---|
| | $^{116}_{52}Te$ | $^{150}_{64}Gd$ | $^{152}_{66}Dy$ |
| KLD for Brody distribution related to GUE | 3.0540 | 3.8853 | 4.2270 |
| KLD for new distribution related to GUE | 1.2594 | 1.1796 | 1.6825 |

Table7. KLD for nuclei with U(5) symmetry with total boson number N=8,9,10 respectively related to GUE.

| KLD for different distributions | Nuclei with SO(6) symmetry | | |
|---|---|---|---|
| | $^{192}_{80}Hg$ | $^{130}_{58}Ce$ | $^{188}_{80}Hg$ |
| KLD for Brody distribution related to GOE | 0.8054 | 1.2217 | 0.8405 |
| KLD for new distribution related to GOE | 0.4823 | 0.8526 | 0.7728 |

Table8. KLD for nuclei with SO(6) symmetry with total boson number N=8,9,10 respectively related to GOE.

These results obviously verify better distances of new suggested distribution to both limits in compared to Brody one. Also we have calculated CRLB for these two distributions in order to investigate properties of this new distribution and compare it with Brody distribution (which is the most popular distribution in statistical analysis [13,17-20,35]), As it has been introduced in [45,47], CRLB for vector functions is

$$CRLB: \frac{\partial \rho(\theta)}{\partial \theta^T}[F(\theta)]^{-1}\frac{\partial \rho^T(\theta)}{\partial \theta}\bigg| \text{with final values of } \alpha, \beta, b \text{ and } q \text{ obtained of MLE method} \quad (25)$$

( see Appendix(I) and also Appendix(II) in [45] for more details about calculation of these quantities).

As it has been presented in tables (9), the new distribution has smaller CRLB in compare with Brody distribution in the same sequences.

| CRLB for distributions  Nuclei | Brody distribution | New distribution |
|---|---|---|
| $^{116}_{52}Te$ | $1.54 \times 10^{-5}$ | $2.25 \times 10^{-8}$ |
| $^{120}_{52}Te$ | $5.45 \times 10^{-9}$ | $9.74 \times 10^{-10}$ |
| $^{124}_{54}Xe$ | $1.58 \times 10^{-11}$ | $3.56 \times 10^{-15}$ |
| $^{108}_{46}Pd$ | $9.88 \times 10^{-8}$ | $3.31 \times 10^{-12}$ |
| $^{104}_{44}Ru$ | $1.95 \times 10^{-12}$ | $1.03 \times 10^{-19}$ |
| $^{192}_{78}Pt$ | $3.1 \times 10^{-7}$ | $1.62 \times 10^{-10}$ |
| $^{192}_{80}Hg$ | $7.85 \times 10^{-8}$ | $1.1 \times 10^{-8}$ |

Table9.CRLB for both distribution in sequences of nuclei with N=8.For similarity of results, we don't reply for other nuclei. All values have been calculated with MLE results (in [45],we have displayed smaller bounds for MLE results in compare with fitting results)

We also apply the above mentioned method to three cases N=25,50,100 to control universality behavior of distribution function in the N→ ∞ as similar to [13-14].We have evaluated all energy levels for these cases by the above explained method (the values of Hamiltonian's parameter in every case are displayed in captions of figures 7,8,9).We have constructed our used sequences from only $2^+$ levels. In the case N=25 (which is close to realistic one), similar behavior (as previous results) can be seen, our results show regular behavior for both symmetry limits (($c_s$=0) vibrational and ($c_s$=1) γ-unstable rotor (SO(6)) limits) but in the cases of N=50 and 100 cases, the effect of boson number makes an universality behavior (tend to Poisson limit) for both symmetry limits and also for transitional region. Table (6) presents these results and verifies our prediction about this special behavior for systems in the case $N \to \infty$ .Figures 7,8,9 display this fact and present regular behavior for case $N \to \infty$ independent of $c_s$ values.

| $c_s$ | 0 | 0.50 | 1 |
|---|---|---|---|
| q (for N=25) | 0.36 ± 0.05 | 0.40 ± 0.06 | 0.38 ± 0.06 |
| q (for N=50) | 0.13 ± 0.04 | 0.18 ± 0.06 | 0.20 ± 0.05 |
| q (for N=100) | 0.08 ± 0.02 | 0.05 ± 0.02 | 0.04 ± 0.02 |

Table10. Brody distribution's parameter for unrealistic cases N=25,50,100 in three regions, $c_s = 0$ indicates U(5) limit, $c_s = 0.5$ presents transitional region and $c_s = 1$ corresponds to SO(6) limit.

## Summary and Remarks

In summary, we have investigated level statistics of both U(5) and SO(6) dynamical symmetries and also transitional regions between these two limits in the SU(1,1) Algebraic approach to Interaction Boson Model(IBM) .We solved Beth-Ansatz equations within an infinite dimensional Lie Algebra. For this, we apply Least square fitting to experimental data (nuclei with these symmetries) and we have evaluated all energy levels below $\leq 7 Mev$ by resultant constants of Hamiltonian .Through the unfolding processes and MLE method, Brody distribution's parameter was obtained that describe regularity or chaotic properties of every spectra. Our results verify theoretical prediction about regular behavior for both limits in compare with transitional regions. Also with controlling the dependence of Brody distribution's parameter to $c_s$ , we can regard $c_s$ as control parameter ( in this approach ) which display phase transitional behavior (tend to GOE type) in the interval $c_s: 0.4 \sim 0.6$ . In order to control distance of our results to both GUE and GOE, we have suggested a new distribution and have evaluated the parameters of this new distribution. The new suggested distribution, has the least KLD distance in compare with GOE and the previous result about chaotic behavior in transitional region in compared to both symmetry limits, have been verified. Also with KLD, we have displayed closer distance of this new distribution to both GUE and GOE limits in compared to Brody one. We also investigated the statistical behaviors of systems in the case $N \rightarrow \infty$ , which verified previous results and show an universality behavior (tend to Poisson limit) for all cases without any dependence to $c_s$ values.

# Appendix(I)

The new distribution is
$$P(s) = b(1+q)(\alpha s^q + \beta s^{q+1})e^{-bs^{q+1}}, \qquad (I-1)$$

With multiplication of all P(s)'s, we can introduce likelihood function as

$$L(q,b,\alpha,\beta) = \prod_{i=1}^{n} b(1+q)\left(\alpha s_i^q + \beta s_i^{q+1}\right)e^{-bs_i^{q+1}} \qquad (I-2a)$$

Or

$$L(q,b,\alpha,\beta) = (b(1+q))^n \prod_{i=1}^{n}\left(\alpha s_i^q + \beta s_i^{q+1}\right)e^{-bs_i^{q+1}}, \qquad (I-2b)$$

We will use logarithm (AI-2) in order to introduce our estimators for all variables as

$$\ln L(q,b,\alpha,\beta) = n\ln(b(1+q)) + \sum_{i=1}^{n}\ln\left(\alpha s_i^q + \beta s_i^{q+1}\right) - b\sum_{i=1}^{n} s_i^{q+1} \qquad (I-2c)$$

$$\frac{\partial \ln L(q,b,\alpha,\beta)}{\partial q} = 0 \Rightarrow f_1: \frac{n}{1+q} + \sum_{i=1}^{n}\ln s_i - b\sum_{i=1}^{n}\ln s_i \, s_i^{q+1} \quad \text{for } q \qquad (I-3a)$$

$$\frac{\partial \ln L(q,b,\alpha,\beta)}{\partial b} = 0 \Rightarrow f_2: \frac{n}{b} - \sum_{i=1}^{n} s_i^{q+1} \quad \text{for } b \qquad (I-3b)$$

$$\frac{\partial \ln L(q,b,\alpha,\beta)}{\partial \alpha} = 0 \Rightarrow f_3: \sum_{i=1}^{n} \frac{s_i^q}{\alpha s_i^q + \beta s_i^{q+1}} \quad \text{for } \alpha \qquad (I-3c)$$

$$\frac{\partial \ln L(q,b,\alpha,\beta)}{\partial \beta} = 0 \Rightarrow f_4: \sum_{i=1}^{n} \frac{s_i^{q+1}}{\alpha s_i^q + \beta s_i^{q+1}} \quad \text{for } \beta \qquad (I-3d)$$

We must take the derivates of all $f_i$ with related to all four variables to construct our Jacobian matrix for Newton-Raphson iteration method as

$$\frac{\partial f_1}{\partial q} = -\frac{n}{(1+q)^2} - b\sum_{i=1}^{n}(\ln s_i)^2 s_i^{q+1} \qquad (I-4a)$$

$$\frac{\partial f_1}{\partial b} = -\sum_{i=1}^{n}\ln s_i \, s_i^{q+1} \qquad (I-4b)$$

$$\frac{\partial f_1}{\partial \alpha} = 0 \qquad (I-4c)$$

$$\frac{\partial f_1}{\partial \beta} = 0 \qquad (I-4d)$$

And similary, for second estimator

$$\frac{\partial f_2}{\partial q} = -\sum_{i=1}^{n} s_i^{q+1} \ln s_i \qquad (I-5a)$$

$$\frac{\partial f_2}{\partial b} = -\frac{n}{b^2} \qquad (I-5b)$$

$$\frac{\partial f_2}{\partial \alpha} = 0 \qquad (I-5c)$$

$$\frac{\partial f_2}{\partial \beta} = 0 \qquad (I-5d)$$

And for third estimator

$$\frac{\partial f_3}{\partial q} = 0 \qquad (I-6a)$$

$$\frac{\partial f_3}{\partial b} = 0 \qquad (I-6b)$$

$$\frac{\partial f_3}{\partial \alpha} = -\sum_{i=1}^{n} \frac{s_i^{2q}}{\left(\alpha s_i^q + \beta s_i^{q+1}\right)^2} \qquad (I-6c)$$

$$\frac{\partial f_3}{\partial \beta} = -\sum_{i=1}^{n} \frac{s_i^{2q+1}}{\left(\alpha s_i^q + \beta s_i^{q+1}\right)^2} \qquad (I-6d)$$

And for fourth one, we have

$$\frac{\partial f_4}{\partial q} = 0 \qquad (I-7a)$$

$$\frac{\partial f_4}{\partial b} = 0 \qquad (I-7b)$$

$$\frac{\partial f_4}{\partial \alpha} = -\sum_{i=1}^{n} \frac{s_i^{2q+1}}{\left(\alpha s_i^q + \beta s_i^{q+1}\right)^2} \qquad (I-7c)$$

$$\frac{\partial f_4}{\partial \beta} = -\sum_{i=1}^{n} \frac{s_i^{2q+2}}{\left(\alpha s_i^q + \beta s_i^{q+1}\right)^2} \qquad (I-7d)$$

Now, we can apply Newton-Raphson iteration as

$$x_{new}^i = x_{old}^i - Df^{-1}(x_{old}^i) \qquad x^i: q, b, \alpha, \beta \qquad (I-8a)$$

$$\begin{bmatrix} q_{new} \\ b_{new} \\ \alpha_{new} \\ \beta_{new} \end{bmatrix} = \begin{bmatrix} q_{old} \\ b_{old} \\ \alpha_{old} \\ \beta_{old} \end{bmatrix} - Df^{-1}(q_{old}, b_{old}, \alpha_{old}, \beta_{old}) f(q_{old}, b_{old}, \alpha_{old}, \beta_{old}) \qquad (I-8b)$$

With applying these relations to our case, final results in order to evaluate our four parameters are obtained:

$$Denominator: \left[-\sum_{i=1}^{n} \ln s_i \, s_i^{q+1}\right]^2 \left[\sum_{i=1}^{n} \frac{s_i^{2q+1}}{(\alpha s_i^q + \beta s_i^{q+1})^2}\right]^2 -$$

$$- \left[-\frac{n}{(1+q)^2} - b\sum_{i=1}^{n} (\ln s_i)^2 s_i^{q+1}\right] \left[-\frac{n}{b^2}\right] \left[\sum_{i=1}^{n} \frac{s_i^{2q+1}}{(\alpha s_i^q + \beta s_i^{q+1})^2}\right]^2 -$$

$$- \left[-\sum_{i=1}^{n} \ln s_i \, s_i^{q+1}\right]^2 \left[-\sum_{i=1}^{n} \frac{s_i^{2q}}{(\alpha s_i^q + \beta s_i^{q+1})^2}\right] \left[-\sum_{i=1}^{n} \frac{s_i^{2q+2}}{(\alpha s_i^q + \beta s_i^{q+1})^2}\right] +$$

$$+ \left[-\frac{n}{(1+q)^2} - b\sum_{i=1}^{n} (\ln s_i)^2 s_i^{q+1}\right] \left[-\frac{n}{b^2}\right] \left[-\sum_{i=1}^{n} \frac{s_i^{2q}}{(\alpha s_i^q + \beta s_i^{q+1})^2}\right] \left[-\sum_{i=1}^{n} \frac{s_i^{2q+2}}{(\alpha s_i^q + \beta s_i^{q+1})^2}\right] \qquad (I-9)$$

$$q_{new} = q_{old} - \left\{ \frac{-\left[-\sum_{i=1}^{n} \frac{s_i^{2q+1}}{(\alpha s_i^q + \beta s_i^{q+1})^2}\right]^2 \left[-\frac{n}{b^2}\right]}{Denominator} + \right.$$

$$+ \frac{\left[-\sum_{i=1}^{n} \frac{s_i^{2q+2}}{(\alpha s_i^q + \beta s_i^{q+1})^2}\right] \left[-\sum_{i=1}^{n} \frac{s_i^{2q}}{(\alpha s_i^q + \beta s_i^{q+1})^2}\right] \left[-\frac{n}{b^2}\right]}{Denominator} \right\} \times$$

$$\times \left\{ \frac{n}{1+q} + \sum_{i=1}^{n} \ln s_i - b \sum_{i=1}^{n} \ln s_i \, s_i^{q+1} \right\} +$$

$$+ \left\{ \frac{\left[-\sum_{i=1}^{n} \frac{s_i^{2q+1}}{(\alpha s_i^q + \beta s_i^{q+1})^2}\right]^2 \left[-\sum_{i=1}^{n} \ln s_i \, s_i^{q+1}\right]}{Denominator} - \right.$$

$$- \frac{\left[-\sum_{i=1}^{n} \ln s_i \, s_i^{q+1}\right] \left[-\sum_{i=1}^{n} \frac{s_i^{2q+2}}{(\alpha s_i^q + \beta s_i^{q+1})^2}\right] \left[-\sum_{i=1}^{n} \frac{s_i^{2q}}{(\alpha s_i^q + \beta s_i^{q+1})^2}\right]}{Denominator} \right\} \times \left\{ \frac{n}{b} - \sum_{i=1}^{n} s_i^{q+1} \right\} \qquad (I-10)$$

$$b_{new} = b_{old} - \left\{ \frac{\left[-\sum_{i=1}^{n} \frac{s_i^{2q+1}}{(\alpha s_i^q + \beta s_i^{q+1})^2}\right]^2 \left[-\sum_{i=1}^{n} \ln s_i \, s_i^{q+1}\right]}{\text{Denominator}} - \right.$$

$$\left. - \frac{\left[-\sum_{i=1}^{n} \frac{s_i^{2q+2}}{(\alpha s_i^q + \beta s_i^{q+1})^2}\right]\left[-\sum_{i=1}^{n} \frac{s_i^{2q}}{(\alpha s_i^q + \beta s_i^{q+1})^2}\right]\left[-\sum_{i=1}^{n} \ln s_i \, s_i^{q+1}\right]}{\text{Denominator}} \right\} \times$$

$$\times \left\{ \frac{n}{1+q} + \sum_{i=1}^{n} \ln s_i - b \sum_{i=1}^{n} \ln s_i \, s_i^{q+1} \right\} +$$

$$+ \left\{ \frac{-\left[-\sum_{i=1}^{n} \frac{s_i^{2q+1}}{(\alpha s_i^q + \beta s_i^{q+1})^2}\right]^2 \left[-\frac{n}{(1+q)^2} - b \sum_{i=1}^{n} (\ln s_i)^2 s_i^{q+1}\right]}{\text{Denominator}} + \right.$$

$$\left. + \frac{\left[-\sum_{i=1}^{n} \frac{s_i^{2q+2}}{(\alpha s_i^q + \beta s_i^{q+1})^2}\right]\left[-\sum_{i=1}^{n} \frac{s_i^{2q}}{(\alpha s_i^q + \beta s_i^{q+1})^2}\right]\left[-\frac{n}{(1+q)^2} - b \sum_{i=1}^{n} (\ln s_i)^2 s_i^{q+1}\right]}{\text{Denominator}} \right\} \times$$

$$\times \left\{ \frac{n}{b} - \sum_{i=1}^{n} s_i^{q+1} \right\} \qquad (I-11)$$

$$\alpha_{new} = \alpha_{old} - \left\{ \frac{-\left[-\sum_{i=1}^{n} \ln s_i \, s_i^{q+1}\right]^2 \left[-\sum_{i=1}^{n} \frac{s_i^{2q+2}}{(\alpha s_i^q + \beta s_i^{q+1})^2}\right]}{\text{Denominator}} + \right.$$

$$\left. + \frac{\left[-\frac{n}{(1+q)^2} - b \sum_{i=1}^{n} (\ln s_i)^2 s_i^{q+1}\right]\left[-\sum_{i=1}^{n} \frac{s_i^{2q+2}}{(\alpha s_i^q + \beta s_i^{q+1})^2}\right]\left[-\frac{n}{b^2}\right]}{\text{Denominator}} \right\} \times \left\{ \sum_{i=1}^{n} \frac{s_i^q}{\alpha s_i^q + \beta s_i^{q+1}} \right\} +$$

$$+ \left\{ \frac{\left[-\sum_{i=1}^{n} \ln s_i \, s_i^{q+1}\right]^2 \left[-\sum_{i=1}^{n} \frac{s_i^{2q+1}}{(\alpha s_i^q + \beta s_i^{q+1})^2}\right]}{\text{Denominator}} - \right.$$

$$\left. - \frac{\left[-\frac{n}{(1+q)^2} - b \sum_{i=1}^{n} (\ln s_i)^2 s_i^{q+1}\right]\left[-\frac{n}{b^2}\right]\left[-\sum_{i=1}^{n} \frac{s_i^{2q+1}}{(\alpha s_i^q + \beta s_i^{q+1})^2}\right]}{\text{Denominator}} \right\} \times \left\{ \sum_{i=1}^{n} \frac{s_i^{q+1}}{\alpha s_i^q + \beta s_i^{q+1}} \right\} \quad (I-12)$$

$$\beta_{new} = \beta_{old} - \left\{ \frac{\left[-\sum_{i=1}^{n} \ln s_i \, s_i^{q+1}\right]^2 \left[-\sum_{i=1}^{n} \frac{s_i^{2q+1}}{(\alpha s_i^q + \beta s_i^{q+1})^2}\right]}{\text{Denominator}} - \right.$$

$$\left. - \frac{\left[-\sum_{i=1}^{n} \frac{s_i^{2q+1}}{(\alpha s_i^q + \beta s_i^{q+1})^2}\right]\left[-\frac{n}{(1+q)^2} - b \sum_{i=1}^{n} (\ln s_i)^2 s_i^{q+1}\right]\left[-\frac{n}{b^2}\right]}{\text{Denominator}} \right\} \times \left\{ \sum_{i=1}^{n} \frac{s_i^q}{\alpha s_i^q + \beta s_i^{q+1}} \right\} +$$

$$+\left\{\frac{-[-\sum_{i=1}^{n}\ln s_i \, s_i^{q+1}]^2 \left[-\sum_{i=1}^{n}\frac{s_i^{2q}}{(\alpha s_i^q + \beta s_i^{q+1})^2}\right]}{\text{Denominator}}+\right.$$

$$\left.+\frac{\left[-\frac{n}{(1+q)^2}-b\sum_{i=1}^{n}(\ln s_i)^2 s_i^{q+1}\right]\left[-\frac{n}{b^2}\right]\left[-\sum_{i=1}^{n}\frac{s_i^{2q}}{(\alpha s_i^q + \beta s_i^{q+1})^2}\right]}{\text{Denominator}}\right\} \times \left\{\sum_{i=1}^{n}\frac{s_i^{q+1}}{\alpha s_i^q + \beta s_i^{q+1}}\right\} \quad (I-13)$$

In order to evaluate a appropriate result for these parameters, as have explained in [45] and also applied for Brody distribution, we have began our calculation with fitting results and with applying iteration processes with program designed in MATLAB software, we can evaluate final results for these parameters with enough accuracy. We have evaluated our calculations with $\alpha$ and $\beta$ independent of "q" and "b", but as have displayed in figure.10.any variation dosen't occur in iteration stages from relation (19) for these quantities.

# Appendix(II)

## CRLB for new distribution

As have explained in [45], we must use the vector form of CRLB as have introduced in [47]

$$cov_\theta(T(X)) \geq \frac{\partial \rho(\theta)}{\partial \theta^T}[F(\theta)]^{-1}\frac{\partial \rho^T(\theta)}{\partial \theta}, \qquad (II-1)$$

The CRLB for distribution will be as

CRLB: $\frac{\partial \rho(\theta)}{\partial \theta^T}[F(\theta)]^{-1}\frac{\partial \rho^T(\theta)}{\partial \theta}\Big|_{with\ final\ values\ of\ \alpha,\beta,b\ and\ q\ obtained\ of\ MLE\ method}$, $(II-2)$

Now for our distribution, we have

$$\theta_1 \to q, \theta_2 \to b$$

And

$$\rho_1 \to \frac{1}{1+q} \Rightarrow \frac{\partial \rho_1}{\partial q} = \frac{-1}{(1+q)^2}, \quad \frac{\partial \rho_1}{\partial b} = 0 \quad \& \quad \rho_2 \to \frac{1}{b} \Rightarrow \frac{\partial \rho_2}{\partial q} = 0, \quad \frac{\partial \rho_2}{\partial q} = \frac{-1}{b^2}, \qquad (II-3)$$

On the other hand, for Fisher integral

$$F(\theta) = \begin{bmatrix} E\left[(X_q - \bar{X}_q)^2\right] & E[(X_q - \bar{X}_q)(X_b - \bar{X}_b)] \\ E[(X_q - \bar{X}_q)(X_b - \bar{X}_b)] & E[(X_b - \bar{X}_b)^2] \end{bmatrix}, \qquad (II-4)$$

Where

$$X_q = \frac{\partial \ln L(q,b)}{\partial q} = \frac{n}{1+q} + \sum_{i=1}^{n} \ln s_i - b\sum_{i=1}^{n} \ln s_i\ s_i^{q+1} \quad \& \quad \bar{X}_q = \frac{1}{n}\sum X_q \qquad (II-5)$$

$$X_b = \frac{\partial \ln L(q,b)}{\partial b} = \frac{n}{b} - \sum_{i=1}^{n} s_i^{q+1} \quad \& \quad \bar{X}_b = \frac{1}{n}\sum X_b \qquad (II-6)$$

Which can combine to final form

$$CRLB: \begin{pmatrix} \frac{\partial \rho_1}{\partial q} & \frac{\partial \rho_1}{\partial b} \\ \frac{\partial \rho_2}{\partial q} & \frac{\partial \rho_2}{\partial q} \end{pmatrix} \begin{bmatrix} E\left[(X_q - \bar{X}_q)^2\right] & E[(X_q - \bar{X}_q)(X_b - \bar{X}_b)] \\ E[(X_q - \bar{X}_q)(X_b - \bar{X}_b)] & E[(X_b - \bar{X}_b)^2] \end{bmatrix}^{-1} \begin{pmatrix} \frac{\partial \rho_1}{\partial q} & \frac{\partial \rho_2}{\partial q} \\ \frac{\partial \rho_1}{\partial b} & \frac{\partial \rho_2}{\partial q} \end{pmatrix} \qquad (II-7)$$

**Or**

$$CRLB: \frac{[E\left[(X_q - \bar{X}_q)^2\right]E[(X_b - \bar{X}_b)^2] - (E[(X_q - \bar{X}_q)(X_b - \bar{X}_b)])^2]}{b^4(1+q)^4(E\left[(X_q - \bar{X}_q)^2\right]E[(X_b - \bar{X}_b)^2] - (E[(X_q - \bar{X}_q)(X_b - \bar{X}_b)])^2)} \qquad (II-8)$$

# Figure caption

Figure1(coloronline).P(s) histograms (NNSDs) for different nuclei with total boson number N=8,the constants of Hamiltonian related to every nuclei was presented in table(1).In all of these NNSDs. horizantal axis represent level spacing (s) and vertical one display P(s), in all figures, the solid lines and dashed lines describe the GOE and Poisson statistics, respectively.

Figure2(coloronline).P(s) histograms (NNSDs) for different nuclei with total boson number N=9,the constants of Hamiltonian related to every nuclei was presented in table(2). In all of these NNSDs. horizantal axis represent level spacing (s) and vertical one display P(s), in all figures, the solid lines and dashed lines describe the GOE and Poisson statistics, respectively.

Figure3(coloronline).P(s) histograms (NNSDs) for different nuclei with total boson number N=10,the constants of Hamiltonian related to every nuclei was presented in table(3). In all of these NNSDs. horizantal axis represent level spacing (s) and vertical one display P(s), in all figures, the solid lines and dashed lines describe the GOE and Poisson statistics, respectively.

Figure4(coloronline). CRLB for estimation process related to(from left to right) N=8 ($^{192}_{78}Pt$ nuclei),N=9($^{114}_{48}Cd$ nuclei) and N=10( $^{120}_{54}Xe$ nuclei) respectively.Other curves don't reply beacuse similarity. In these graphs, the horizontal axis represents number of iteration and vertical one, represents $var(f) - \frac{1}{M\,F(q)}$.

Figure5(coloronline). Relation between "q"(vertical axis) and $c_s$ (horizontal one). The maximum values of "q" occur in the region $c_s \sim 0.4 \to \sim 0.6$ for N=8.9.10 respectively.

Figure6(coloronline). NNSDs for both symmetry limits (U(5) and SO(6)) and transitional region.Used sequences was prepared from experimental data[48-54]related to $2^+$ to $4^+$ levels of nuclei was introduced in tables (1-3). Our criterion factor in classification was the values of $c_s$ .Nuclei with $c_s \sim 0 \to 0.2$ was grouped in U(5) limit, Nuclei with $c_s \sim 0.8 \to 1$ was grouped in SO(6) limit, and Nuclei with $c_s \sim 0.4 \to 0.6$ was grouped in transitional region, in all figures, the solid lines and dashed lines describe the GOE and Poisson statistics, respectively.

Figure7(coloronline). NNSDs for unrealistic system with N=25 . our used specta was constructed with $\alpha = 300 kev, g = 1 kev, \gamma = 95.2166 kev, \delta = -11.2466 kev, c_d = 1$ and $c_s = 0, 0.5\ and\ 1$ respectively., in all figures, the solid lines and dashed lines describe the GOE and Poisson statistics, respectively. We don't reply our calculation with other values of these parameters as[13] because any significant changes don't ocuur

Figure8(coloronline). NNSDs for unrealistic system with N=50 . our used specta was constructed with $\alpha = 320 kev, g = 1 kev, \gamma = 64.7585 kev, \delta = 3.7727 kev, c_d = 1$ and $c_s = 0, 0.5\ and\ 1$ respectively. in all figures, the solid lines and dashed lines describe the GOE and Poisson statistics, respectively. We don't reply our calculation with other values of these parameters as[13] because any significant changes don't ocuur

Figure9(coloronline). NNSDs for unrealistic system with N=100 . our used specta was constructed with $\alpha = 310 kev, g = 1 kev, \gamma = 80.2491 kev, \delta = 11.317 kev, c_d = 1$ and $c_s = 0, 0.5\ and\ 1$ respectively. in all figures, the solid lines and dashed lines describe the GOE and Poisson statistics, respectively.We don't reply our calculation with other values of these parameters as[13] because any significant changes don't ocuur

Figure10. (color online).variation of our proposed constant for new distribution in different iteration stages which verify our aim that any change dose'nt occur with the main distribution .The left one represented for $\alpha$ which horizontal axis represent number of iteration and vertical one represent $A(\equiv \alpha \Big/ 1 - \dfrac{\left(\dfrac{\Gamma[\frac{q+2}{q+1}]}{b^{\frac{1}{1+q}}}\right)^2 - \dfrac{\Gamma[\frac{q+2}{q+1}]}{b^{\frac{1}{1+q}}}}{\left(\dfrac{\Gamma[\frac{q+2}{q+1}]}{b^{\frac{1}{1+q}}}\right)^2 - \dfrac{\Gamma[\frac{q+3}{q+1}]}{b^{\frac{2}{1+q}}}})$ ) and the right one display variation of $\beta$ which the horizontal axis represent number of iteration and vertical one represent $B(\equiv \beta \Big/ \dfrac{\left(\dfrac{\Gamma[\frac{q+2}{q+1}]}{b^{\frac{1}{1+q}}}\right) - 1}{\left(\dfrac{\Gamma[\frac{q+2}{q+1}]}{b^{\frac{1}{1+q}}}\right)^2 - \dfrac{\Gamma[\frac{q+3}{q+1}]}{b^{\frac{2}{1+q}}}})$ ).

Figure1.

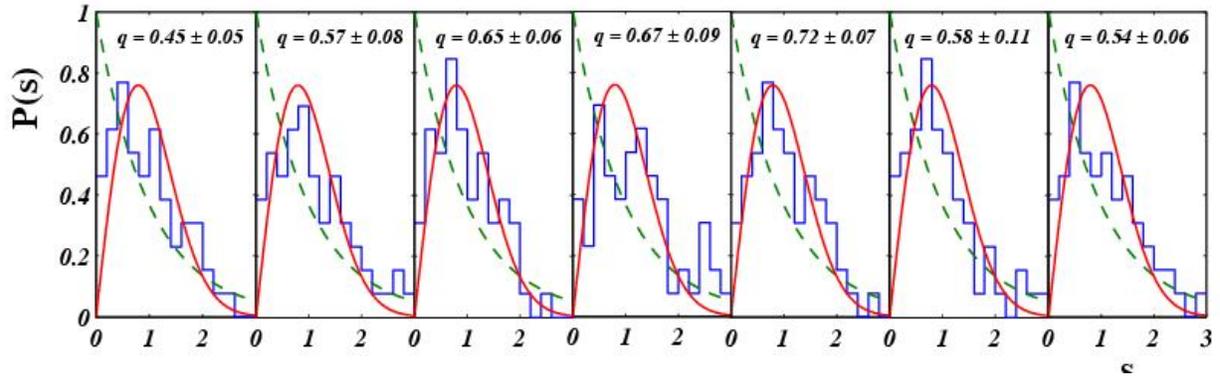

Figure2.

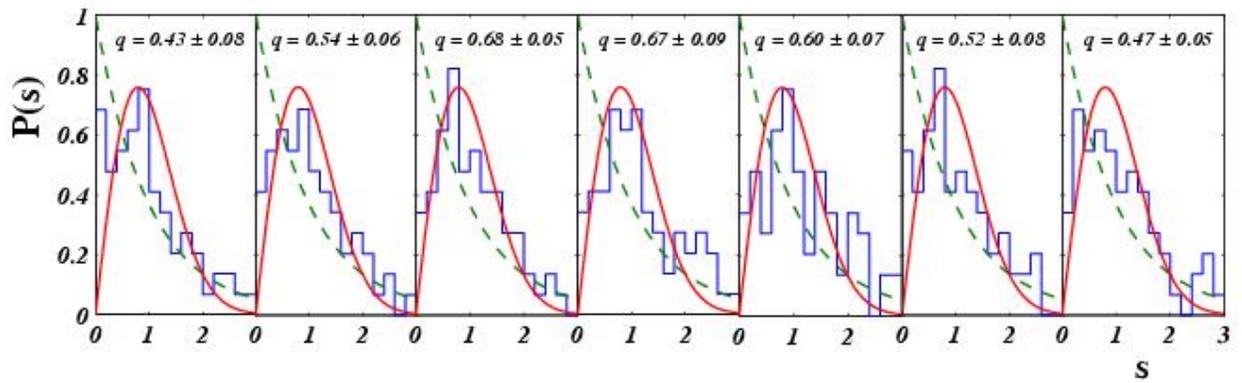

Figure3.

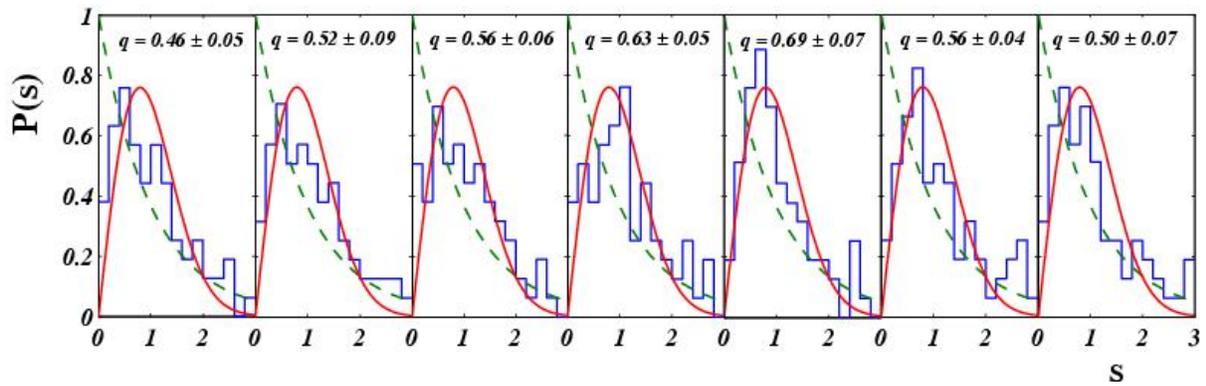

Figure4.

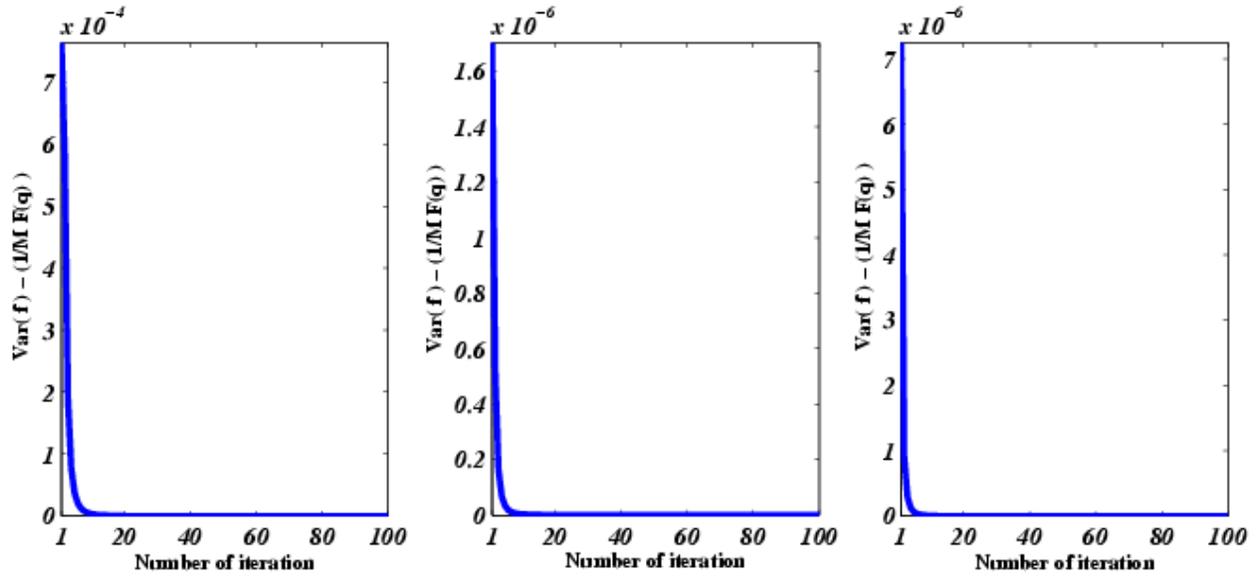

Figure5.

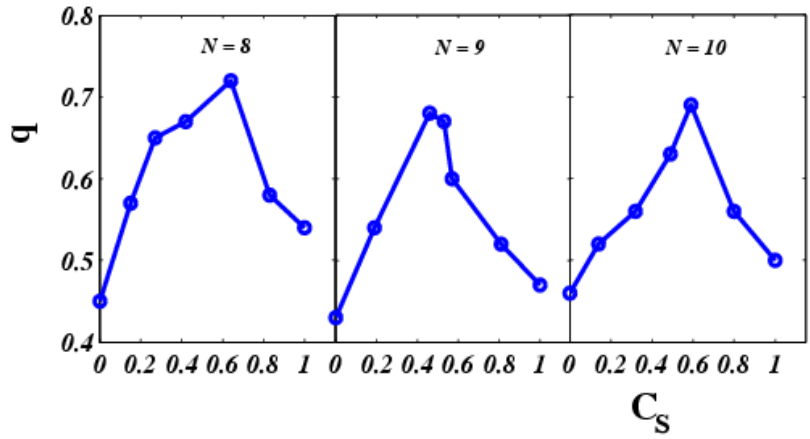

Figure6.

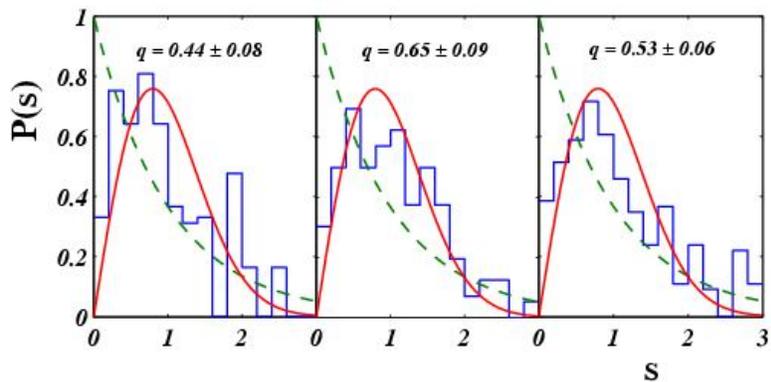

Figurre7.

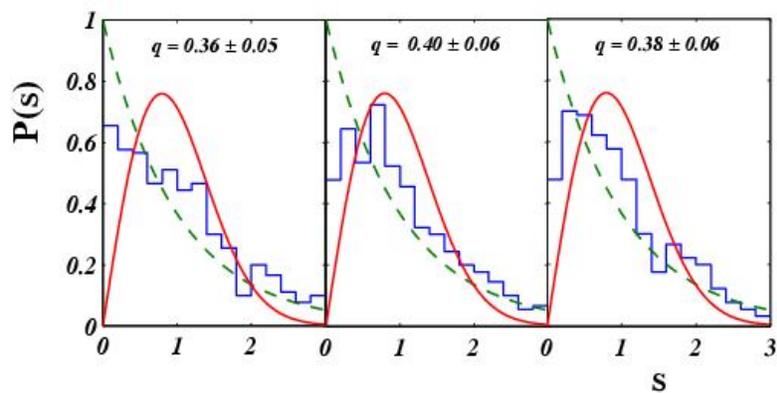

Figure8.

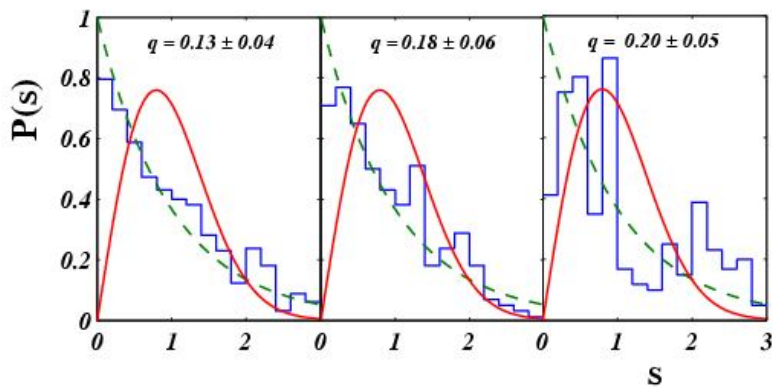

Figure9.

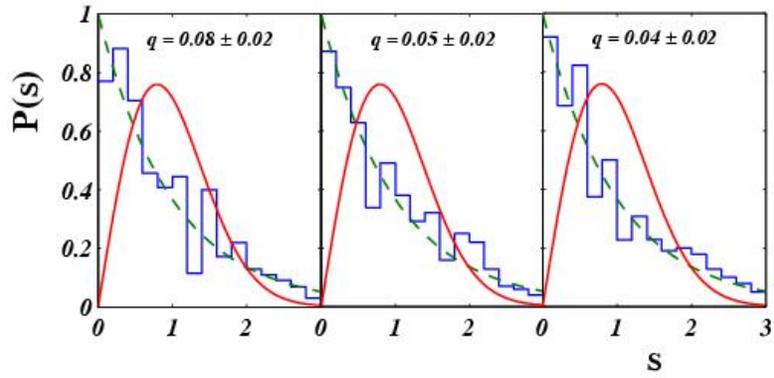

Figure10.

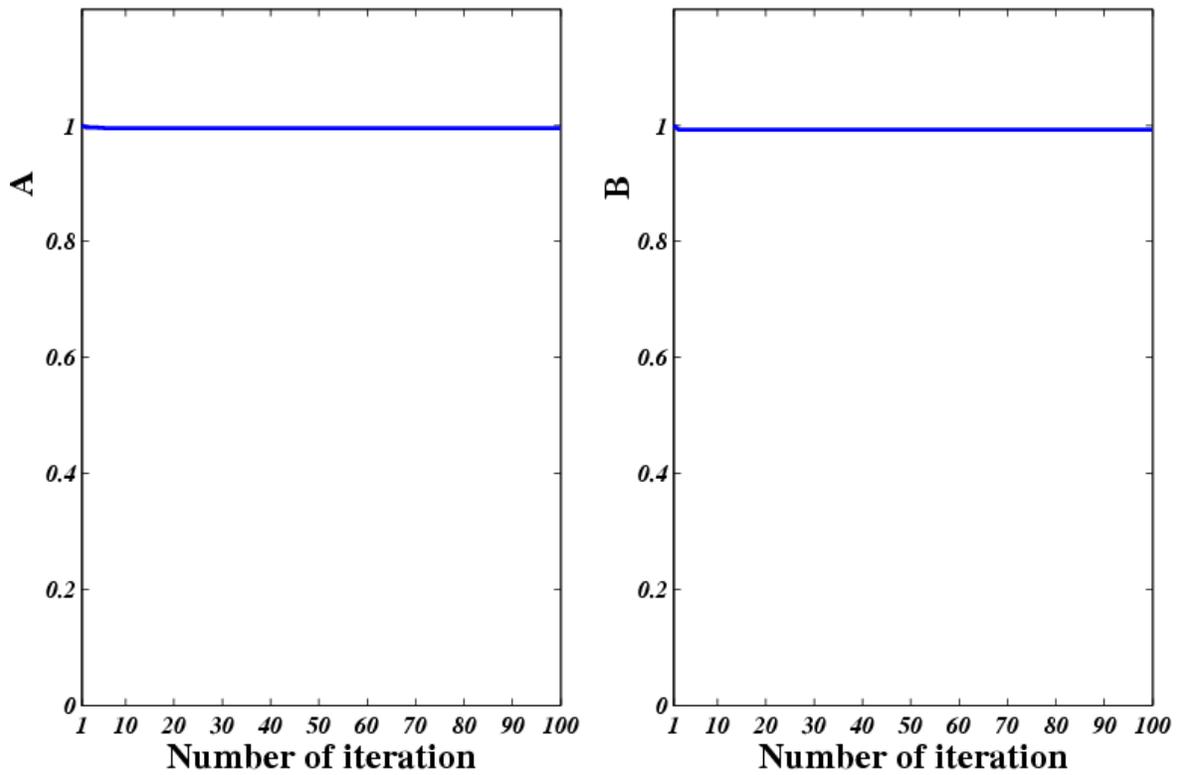